\documentclass[prd,preprintnumbers,floatfix,aps,nofootinbib,notitlepage]{revtex4}

\usepackage{latexsym}
\usepackage{epsfig}
\usepackage{amssymb}
\usepackage{amsmath}
\usepackage{dcolumn}
\usepackage{bm}
\usepackage{multirow}
\usepackage{color}

\newcommand{\lp}{\left(}
\newcommand{\rp}{\right)}
\newcommand{\lb}{\left[}
\newcommand{\rb}{\right]}

\newcommand{\ba}{\begin{eqnarray}}
\newcommand{\ea}{\end{eqnarray}}
\newcommand{\be}{\begin{equation}}
\newcommand{\ee}{\end{equation}}

\newcommand{\al}{\alpha}
\newcommand{\bt}{\beta}

\newcommand{\nab}{{}^{(3)}\nabla}
\newcommand{\ric}{^{(3)}R}
\newcommand{\R}{\mathcal{R}}

\newcommand{\diff}{{\rm d}}
\newcommand{\vp}{Q^{(\phi)}}
\newcommand{\vP}{Q^{(\psi)}}

\begin{document}

\title{A note on viability of nonminimally coupled $f(R)$ theory}

\author{Tomi S. Koivisto}\email{tomi.koivisto@fys.uio.no}
\affiliation{Nordita, KTH Royal Institute of Technology and Stockholm University, Roslagstullsbacken 23, SE-10691 Stockholm, Sweden}
\author{Nicola Tamanini}\email{nicola.tamanini@cea.fr}
\affiliation{Institut de Physique Th\'eorique, CEA-Saclay, CNRS UMR 3681, Universit\'e Paris-Saclay, F-91191 Gif-sur-Yvette, France}

\date{\today}

\begin{abstract}

Consistency conditions for nonminimally coupled $f(R)$ theories have been derived by requiring the absence of tachyons and instabilities in the scalar fluctuations. 
This note confirms these results and clarifies a subtlety regarding different definitions of sound speeds. 

\end{abstract}

\preprint{NORDITA-2016-54}

\maketitle

A class of nonminimally coupled $f(R)$ models can be defined by the action \cite{Koivisto:2005yk,Nojiri:2004bi,Allemandi:2005qs}
\be \label{action1}
S = \int \diff^4x \sqrt{-\hat{g}}\lb f_1(\hat{R}) + f_2(\hat{R})\mathcal{\hat{L}}\rb\,,
\ee
where $\hat{\mathcal{L}}$ is the matter Lagrangian and $\hat{R}$ the Ricci curvature of the metric $\hat{g}_{\mu\nu}$. There has been some interest in such models, especially due to their possible cosmological applications  \cite{Harko:2014gwa}. We have previously considered the fluctuation spectrum of these models in the case of a massless scalar field $\psi$ with the Lagrangian $\hat{\mathcal{L}}=\hat{X}=-\frac{1}{2}\hat{g}^{\al\bt}\psi_{,\al}\psi_{,\bt}$, in order to deduce the conditions that follow by requiring the absence of ghosts, classical instabilities and superluminal propagation of the fluctuations \cite{Tamanini:2013aca}.
These conditions can be regarded as restrictions on the forms of the functions $f_1(\hat{R})$ and $f_2(\hat{R})$ in any viable theory, if one considers that such should be able to accommodate at least a massless scalar field without the aforementioned pathologies. 

The latter point of view was not taken in the recently published paper  \cite{Bertolami:2014hra}, where it was argued e.g.~that the constraints do not apply to ``models in which there is only baryonic matter'' and that ``faster-than-light propagation of perturbations for a particular matter species does not invalidate the underlying theory, but instead shows that the adopted matter content is unphysical''. Although it should not be difficult to see that fermionic instead of scalar, or otherwise more complicated matter Lagrangians than $\hat{\mathcal{L}}=\hat{X}$ in (\ref{action}), would generically result in similar pathologies, here we will not enter into a discussion of potential viability of models that do not admit scalar matter. Ref.~\cite{Bertolami:2014hra} was, however, further concerned that the constraints  presented in Ref.~\cite{Tamanini:2013aca} might not be applicable even in the case $\hat{\mathcal{L}}=\hat{X}=-\frac{1}{2}\hat{g}^{\al\bt}\psi_{,\al}\psi_{,\bt}$. The purpose of this note is to clarify the issues that were raised in Ref.~\cite{Bertolami:2014hra}, detailing some steps omitted in the original derivations of Ref.~\cite{Tamanini:2013aca}. 
\newline

Let us begin by quoting the action (\ref{action1}) rewritten in the Einstein frame (which we here denote with the ``unhatted'' geometric variables) as the following two-scalar theory (cf.~(15) of \cite{Tamanini:2013aca} and (27) of \cite{Bertolami:2014hra}) :
\be \label{action}
S = \int \diff^4x \sqrt{-g}\lb \frac{1}{2\kappa^2} R + Y + p(X,\phi)\rb\,,
\ee
where $Y=-\frac{1}{2}g^{\al\bt}\phi_{,\al}\phi_{,\bt}$ is the kinetic term of the scalar field $\phi$, and the function $p(X,\phi)$, which in general can be very nonlinear, is determined by the two functions in the original action (\ref{action1}). The gravitational coupling is given by $\kappa=1/\sqrt{8\pi G}$, where $G$ is the Newton's constant. The matter kinetic term $X$ is written in the Einstein frame, since the $\phi$-dependence of the rescaled kinetic term can be taken into account by a simple redefinition of the function $p$, to wit $\hat{p}(\hat{X},\phi) =  \hat{p}(e^{-\sqrt{\frac{2}{3}}\kappa\phi}X,\phi) \equiv p(X,\phi)$, without loss of generality. 
That this redefinition is always possible renders the concerns of  Ref.~\cite{Bertolami:2014hra} moot, in particular we do not find a ``more complex dependency of the putative k-essence function $p$ on $\phi$'' that is ``much more evolved than assumed''. In the Appendix (\ref{action1}) we explicitly derive, from the action (\ref{action1}), the equivalent action (\ref{action}).

The remaining main points of Ref.~\cite{Bertolami:2014hra}, to be addressed here, are related to the derivation of the sound speeds for the two fields present in (\ref{action}).
 The misunderstanding of Ref.~\cite{Bertolami:2014hra} is 
 that Ref.~\cite{Tamanini:2013aca} had somehow assumed in the beginning an equivalence with simpler k-essence models, instead of arriving at formally analogous results by rigorous derivation in the end\footnote{The propagation of the two fields in the action \eqref{action} could actually be deduced by just treating the two scalars as test fields and expanding around a flat space, and it is easy to see that only the field $\psi$ has a nontrivial sound speed, stemming from the nonlinear dependence on its kinetic term. In Ref.~\cite{Tamanini:2013aca} we however used the full machinery of the cosmological perturbation theory, perturbing both fields and the metric, to arrive at the results that indeed, at the relevant limit, are compatible with the simplest guess one would base on the ample previous literature on scalar fields in cosmology.}. According to Ref.~\cite{Bertolami:2014hra}, the perturbation analysis resorted to an ``improper comparison'' due to ``an incorrect analogy with $k$-essence'', in particular by neglecting the potentially important effect of the fluctuations in the field $\phi$.

Let us therefore redo the derivation, in an alternative formalism and now keeping specifically track of fluctuations of the field $\phi$ and of the metric $g_{\mu\nu}$, in addition to the fluctuations in the massless, non-minimally coupled scalar $\psi$. Following the calculations of Langlois {\it et al} \cite{Langlois:2008qf}, we first write the metric in the ADM form \cite{Arnowitt:1962hi},
\be
\diff s^2= -N^2\diff t^2 + h_{ij}\lp \diff x^i + N^i \diff t\rp\lp \diff x^j + N^j \diff t\rp\,,
\ee
where $N$ is the lapse and $N^i$ the shift. We will denote the determinant of the spatial metric  $h_{ij}$ by $h$, its associate covariant derivative by $\nab$ and its scalar curvature by $\ric$. The extrinsic curvature of the spatial hypersurfaces is characterised by the symmetric tensor
\be
E_{ij}= \frac{1}{2}\dot{h}_{ij}-\nab_{(i}N_{j)}\,.
\ee
With these definitions, the action (\ref{action}) can be written as
\be \label{adm}
S = \frac{1}{2}\int \diff t \diff^3 x \sqrt{h}\lb \frac{1}{\kappa^2} N \ric + \frac{1}{N\kappa^2}\lp -E^2 + E_{ij}E^{ij}\rp + 2N  Y + 2N p(X,\phi) \rb\,,
\ee 
where the kinetic terms are given by
\be
X   =   \frac{1}{2N^2}\lp \dot{\psi}-N^i \psi_{,i}\rp^2 - \frac{1}{2}h^{ij}\psi_{,i}\psi_{,j}\,, \quad
Y   =   \frac{1}{2N^2}\lp \dot{\phi}-N^i \phi_{,i}\rp^2 - \frac{1}{2}h^{ij}\phi_{,i}\phi_{,j}\,.
\ee
We are interested in finding the spectrum of scalar fluctuations. For convenience, we fix the gauge in the following to the spatially flat one, so we have that $h_{ij}=a^2(t)\delta_{ij}$ and $\ric=0$, and introduce fluctuations around the background values of the metric and the two fields as:
\be \label{expansion}
N=1+\delta N\,, \quad N_i = \partial_i \varphi\,, \quad \phi \rightarrow \phi + \vp\,, \quad \psi \rightarrow \psi + \vP\,.
\ee
It might be useful to note that the standard Bardeen potential $\Psi_B$ and the standard comoving curvature perturbation $\mathcal{R}$ in cosmological perturbation theory are related \cite{Mukhanov:1990me}, with our spatially flat gauge choice, to the shift perturbation $\varphi$ and the scalar field perturbations, respectively,  as
\be \label{comoving}
\Psi_B=-H\varphi\,, \quad \mathcal{R}=\frac{H\lp \dot{\phi}\vp + p_{,X}\dot{\psi}\vP\rp}{2\lp Y + p_{,X}X\rp}\,,
\ee
where $H=\dot{a}/a$ can be identified with the Hubble expansion rate.
The momentum constraint at the zeroth order is satisfied identically, but at the linear order gives
\be \label{constraint}
\delta N = \frac{\kappa^2}{2H}\lp \dot\phi Q^{(\phi)} + p_{,X} \dot\psi Q^{(\psi)} \rp\,.
\ee
When we then insert the expansion (\ref{expansion}) into the action (\ref{adm}) and implement the constraint (\ref{constraint}), it turns out that the contributions from the shift perturbation cancel. The final result, up to second order in perturbations, can be written in terms of a field doublet $\bar{Q} \equiv \lp \vp, \vP\rp$ and four $2\times 2$ matrices as
\be \label{action2}
S =  \frac{1}{2} \int \diff t \diff^3 x a^3 \lb \dot{\bar{Q}} \mathrm{T} \dot{\bar{Q}}^T - h^{ij}\nab_i\bar{Q}\mathrm{K}\nab_j\bar{Q}^T  - \bar{Q}\mathrm{M}\bar{Q}^T + 2\dot{\bar{Q}}\mathrm{S}\bar{Q}^T\rb\,.
\ee
The kinetic matrices are 
\be
\mathrm{T} = 
\left( \begin{array}{cc}
1 & 0  \\
0 & p_{,X} + 2p_{,XX}X  
\end{array} \right)\,, \quad
\mathrm{K} = 
\left( \begin{array}{cc}
1 & 0  \\
0 & p_{,X}  
\end{array} \right)\,.
\ee
These determine the relevant properties of the propagation of the scalar perturbations. The dynamical evolution of the large-wavelength perturbation modes can also be affected by the effective mass term and the mixing term, which, for completeness, are given by
\be
\mathrm{M} = 
\left( \begin{array}{cc}
\mathrm{M}^{(\phi\phi)}  & \frac{p_{,X}}{H}\dot{\psi}\lp 2p_{,X}X-p \rp_{,\phi}  \\
 \frac{p_{,X}}{H}\dot{\psi}\lp 2p_{,X}X-p \rp_{,\phi}    & \mathrm{M}^{(\psi\psi)} 
\end{array} \right)\,, \quad
	\mathrm{S} = 
\left( \begin{array}{cc}
0 & \dot{\psi}p_{,X \phi}  \\
0 & -\frac{2}{H} p_{,X} p_{,XX} X^2  
\end{array} \right)\,,
\ee
where 
\ba
\mathrm{M}^{(\phi\phi)} & = & \frac{\dot{\phi}}{H}\lp 2p_{,X}X-p\rp_{,\phi} -\frac{Y}{H^2}\lp Y+p_{,X}X+2p_{,XX}X^2\rp - \frac{1}{a^3}\frac{\diff}{\diff t}\lp \frac{a^3}{H}Y\rp 
 + 3Y -p_{,\phi\phi}\,, \nonumber \\
 \mathrm{M}^{(\psi\psi)} & = & 3Xp_{,X}^2 - \frac{p_{,X}}{H^2}X\lp Y + p_{,X}X+2p_{,XX}X^2\rp - \frac{1}{a^3}\frac{\diff}{\diff t}\lp \frac{a^3}{H}p_{,X}^2X\rp\,. 
\ea
The sound speeds of the two scalar modes are clearly the eigenvalues of the matrix $\mathrm{T}^{-1}\mathrm{K}$. We obtain 
\be \label{sounds}
c_\phi^2 = 1\,, \quad c_\psi^2 = \frac{p_{,X}}{2Xp_{,XX}+p_{,X}}\,.
\ee 
This is, of course, precisely the expected result \cite{Tamanini:2013aca}. It would not in fact change did we considered $p(\hat{X},\phi)$ instead of $p(X,\phi)$ in action (\ref{action}), since the conformal rescaling cancels out from $c_\psi^2$.  

Let us use this opportunity to clarify different definitions of sound speeds that occur in cosmology.
\begin{itemize}
\item The gauge-invariant propagation velocities of the canonical degrees of freedom; in the case at hand, given as (\ref{sounds}). These determine the evolution of small-wavelength fluctuations \cite{Mukhanov:1990me}. For example, had we $c_\psi^2<0$, it would signal that the field $\psi$ has a gradient instability. 
\item The sound speed squared of a fluid can be also defined as the pressure perturbation divided by the density perturbation evaluted in the rest frame of the fluid. Now, due to the coupling, this definition is slightly ambiguous and naively would give a slightly different result\footnote{One can though check that in practice this does not make a difference for the results as the parameter constraints derived from the two definitions would coincide for all the four explicit example classes of models considered in Ref.~\cite{Tamanini:2013aca}.}  from (\ref{sounds}) because $p(X,\phi)$ contains also perturbations of the field $\phi$.
\item The so called adiabatic sound speed squared is given, for each component, by the ratio of the time derivative of the background density and the time derivative of the background pressure. This coincides with the physical propagation speed of perturbations only if the component has no entropic perturbations, which is not the case for scalar field.
\item The comoving curvature perturbation (\ref{comoving}) is sourced by all the matter content, and thus it is not a surprise that the effective sound speed for $\R$ is given by the collective sound speed of the total matter content, given by a weighted average of the individual sound speeds. 
\end{itemize}

To conclude, the consistency conditions for the models (\ref{action1}) presented in Ref.~\cite{Tamanini:2013aca} are robust. 
We believe they provide useful information and constrain efficiently the allowed parameter space for potentially viable models. For instance, they bring the investigation of the power-law
models $f_2(R) \sim R^n$ to its closure: it was previously known that $0<n<1$ is unstable, and from \cite{Tamanini:2013aca}  we learn that $n<0$ would give rise to superluminal propagation and $n \ge 1$ to a ghost.

\acknowledgements{NT acknowledges support from the Labex P2IO and the Enhanced Eurotalents Programme.}


\section*{Appendix} 
\label{sec:appendix}

In this appendix we rederive the action \eqref{action} from action \eqref{action1}, again using, as a cross-check, a somewhat different method from \cite{Tamanini:2013aca}. There we kept the matter Lagrangian 
$\hat{\mathcal{L}}$ general and applied a Legendre transformation method for both the $R$ and the $\mathcal{\hat{L}}$. Here we specify $\mathcal{\hat{L}}=\hat{X}$ and apply a straightforward
generalisation of a well-known procedure that recasts a minimally coupled $f(R)$ gravity into a scalar-tensor theory.    

So we begin with action (\ref{action1}) where $\mathcal{\hat{L}}=\hat{X}$. Introducing a field $\alpha$ and a Lagrange multiplier $\varphi$ that sets $\alpha=R$ we can write the action equivalently as
\be \label{action3}
S = \int \diff^4x \sqrt{-\hat{g}}\lb f_1(\alpha) + f_2(\alpha)\hat{X} + \varphi\lp \hat{R}-\alpha\rp \rb\,.
\ee
We then vary the action with respect to $\alpha$, and solve it from its own equation of motion
\be
f_1'(\alpha) + f_2'(\alpha)\hat{X}=\varphi \quad \Rightarrow \quad \alpha={A}(\hat{X},\varphi)\,,
\ee
and it is then legitimate to plug the solution back into the action, to obtain
\be
S = \int \diff^4x \sqrt{-\hat{g}}\lb \varphi \hat{R} + 2V(\hat{X},\varphi)\rb\,, \quad \text{where} \quad V(\hat{X},\varphi)=\frac{1}{2}\lp f_1({A})-\varphi {A} + f_2({A})\hat{X}\rp\,.
\ee
Under a conformal rescaling $g_{\mu\nu}=\Omega^2 g_{\mu\nu}$, the relevant quantities transform as
\be
\sqrt{-\hat{g}}=\Omega^{-4}\sqrt{-g}\,, \quad \hat{R}=\Omega^2 R + 6\Omega\Box\Omega-12(\partial\Omega)^2\,, \quad \hat{X}=\Omega^2 X\,.
\ee
Performing the rescaling with the conformal factor 
\be
\Omega^2 = \phi = \sqrt{\frac{3}{2}}\frac{1}{\kappa}\log{\varphi}\,,
\ee
we obtain:
\be
S = \int \diff^4x \sqrt{-g}\lb  \frac{R}{2\kappa^2} + Y + e^{-2\sqrt{\frac{2}{3}}\kappa\phi}\lp f_1(A) - e^{\sqrt{\frac{2}{3}}\kappa\phi}A + e^{\sqrt{-\frac{2}{3}}\kappa\phi}f_2(A)X \rp\rb\,, \quad 
A = A\lp e^{\sqrt{\frac{2}{3}}\kappa\phi},e^{-\sqrt{\frac{2}{3}}\kappa\phi}X\rp\,.
\ee
We have then recovered the same relation between actions \eqref{action1} and \eqref{action} as was obtained in \cite{Tamanini:2013aca}.

\end{document}